\documentclass[letterpaper, 10pt, conference]{IEEEtran}
\IEEEoverridecommandlockouts
\usepackage{nopageno}
\usepackage{cite}
\usepackage{ifpdf}
\usepackage{subfigure}
\usepackage{graphicx}
\DeclareGraphicsExtensions{.pdf,.jpg,.png}
\usepackage[cmex10]{amsmath}
\usepackage{amssymb}
\usepackage{array}
\usepackage{mdwmath}
\usepackage{mdwtab}
\usepackage{pslatex}
\usepackage{xcolor}
\usepackage{textcomp}
\usepackage{placeins}
\usepackage{eqparbox}
\usepackage{url}
\usepackage{multicol}
\usepackage{subfigure}
\usepackage{float}
\usepackage{algorithmicx}
\usepackage{algpseudocode}
\usepackage{amsthm}
\usepackage{comment}
\usepackage{multirow}
\usepackage{mathtools}
\usepackage{setspace}
\def\BibTeX{{\rm B\kern-.05em{\sc i\kern-.025em b}\kern-.08em
		T\kern-.1667em\lower.7ex\hbox{E}\kern-.125emX}}

\DeclarePairedDelimiter\ceil{\lceil}{\rceil}
\DeclarePairedDelimiter\floor{\lfloor}{\rfloor}

\begin{document}
 \pagenumbering{gobble}
\title{Analysis of Uplink Scheduling
	for \\ Haptic Communications}

\author{\IEEEauthorblockN{Maliheh Mahlouji, Toktam Mahmoodi\\}
	\IEEEauthorblockA{Centre for Telecommunications Research, Department of Informatics \\King's College London, London WC2B 4BG, UK}}

\maketitle

\begin{abstract}

While new mechanisms and configurations of the 5G radio are offering step forward in delivery of ultra-reliable low latency communication services in general, and haptic communications in particular, they could inversely impact the remainder of traffic services. In this paper, we investigate the uplink access procedure, how different advances in this procedure enhance delivery of haptic communication, and how it affects the remainder of traffic services in the network. We model this impact as the remainder of service, using stochastic network calculus. Our results show how best the tradeoff between faster or more resource efficient uplink access can be made depending on the rate of haptic data, which is directly relevant to the application domain of haptic communication.


\end{abstract}


\section{Introduction}
\label{sect:intro}

Enabling ultra-reliable low latency communication (URLLC) is one of the main challenges of the next generation mobile networks, a.k.a. 5G. The standardization effort of New Radio (5GNR) has, thus, proposed series of enhancements to reduce the radio transmission delay and step forward in enabling  URLLC for critical applications \cite{3GPPNR}, \cite{Joachim}. However, configurations offering lower latency and higher reliability for the URLLC service, will inversely impact the remainder of traffic services (we refer to this as leftover traffic).

Haptic communication, particularly when used for critical applications, is one the finest examples of the URLLC services, sending haptic data (such as position/velocity, force/torque, texture, etc) from master to slave and receiving the haptic data at the master side. This round-trip communication should be performed within low and stable latency in order for stability of the control scheme to be maintained. There is a rich literature of teleoperation, focusing on balancing the control stability with transparency in order to compensate for communication delay \cite{control}, where control stability is maintained in the expense of lowering the quality of haptic communication. Therefore, communication latency will have the major impact on transparency of the haptic communication and hence will be a deciding factor in whether haptic communication could be used for critical and precise teleoperation \cite{comst-haptic}.

In this paper, we focus on the Uplink (UL) scheduling procedure and the delay introduced through UL scheduling to the communication, as one of the main configuration of 5GNR for URLLC \cite{3GPPNR}, \cite{Joachim}. To this end, we study four different UL scheduling: 1) the dynamic scheduling with scheduling request (SR) procedure for every transmission, as occurs in the 4G-LTE, 2) the semi-persistent scheduling, 3) an adaptation of semi-persistent scheduling that adapts the persistency depending the traffic, namely soft resource reservation, and finally 4) the fast UL, that is considered as an alternative for the 5G. We look into these four scheduling mechanisms, and how employing them for haptic traffic will impact serving the remainder of traffic in the system. While network slicing could potentially provide isolation between these services \cite{icc-slicing}, the way resources are split between different slices is a complex decision for which detailed studies, such as the study in this paper, are essential.

This paper is organized as follows: in section \ref{sec:delay}, the four UL scheduling mechanisms under study are described in details, and the delay they incur to the communication is formulated. In section \ref{sec:SNC}, we present a model to compute radio access delay for the leftover traffic as a result of deploying any of the above mentioned UL scheduling techniques. In section \ref{sec:simulation}, we present the numerical results and finally, in section \ref{sec:conclusions}, concluding remarks are summarized. 

\section{Uplink Scheduling Mechanisms}
\label{sec:delay}
In this section, we will review four potential UL scheduling mechanisms, based on existing scheduling in the LTE, but also potential proposals for ultra-low latency applications. These four scheduling mechanisms are chosen based on the literature (\cite{Joachim}, \cite{ULscheduling}, \cite{3GPPLatency}, \cite{SPS2012}, and \cite{Massimo}) in a way that they cover a diverse range of UL scheduling techniques in terms of their offered delay and communication resource consumption. The main goal of this section is to summarize delay incurred by each of these UL scheduling techniques.

\subsection{Dynamic Scheduling}
In Dynamic Scheduling (DS), user (UE) needs to send Scheduling Request (SR) to eNodeB (eNB) for each packet transmission. This is the UL scheduling technique used in LTE, and it provides significant benefits in terms of spectral efficiency, while also introducing an additional delay due to the involved procedure \cite{ULscheduling}. 

In order to send SR, the UE must wait till the next available SR-valid Physical Uplink Control Channel (PUCCH). In this paper, $T_{SR}$ denotes the period of SR opportunity. Then, eNB will decode the SR and send an UL grant to the UE. After receiving the UL grant, UE decodes the grant and waits for the allocated grant to start transmission over Physical Uplink Shared Channel (PUSCH) \cite{3GPPLatency}.

\begin{figure}
	\centering
	\includegraphics[width=6.5cm]{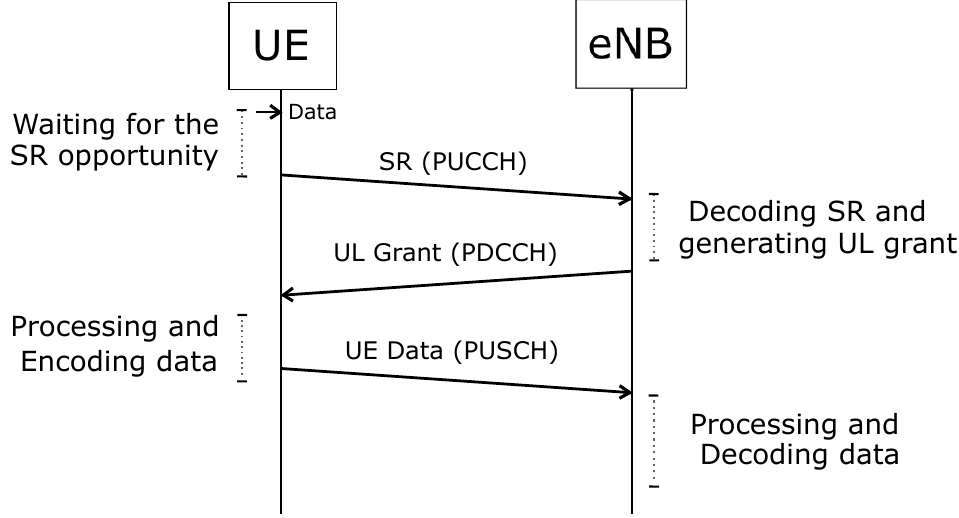}
	\caption{Schematic of Scheduling Request (SR) procedure \cite{3GPPLatency}\vspace{-10mm}}
	\label{fig:SR procedure}
\end{figure}

A schematic of SR procedure in LTE is depicted in Figure \ref{fig:SR procedure}. This procedure repeats for every UL transmission. Therefore, assuming successful signal transmission and reception in all SR procedure, the maximum UL radio access latency for each packet using dynamic scheduling can be expressed as $d_{DS}$, according to Equation (\ref{eq:d_ds}).

\begin{equation}
\begin{aligned}
\label{eq:d_ds}
d_{DS} \leq &T_{SR}+T_{tx(UE \rightarrow eNB)}+T_{proc(eNB)}+T_{tx(eNB \rightarrow UE)}+\\
&T_{proc(UE)}+T_{tx(UE \rightarrow eNB)}+T_{proc(eNB)} \\
=&T_{SR}+TTI+TTI+TTI+ \\
&TTI+TTI+TTI 
=T_{SR}+6TTI \text{,}
\end{aligned}
\end{equation}

where $T_{SR}$ is maximum waiting time for SR opportunity on PUCCH, $T_{tx(UE \rightarrow eNB)}$ and $T_{tx(eNB \rightarrow UE)}$ denote packet transmission time from UE to eNB and from eNB to UE, respectively. $T_{proc(UE)}$ and $T_{proc(eNB)}$ denote processing time for decoding and encoding packets in UE and eNB, respectively. Based on the 3GPP documentation \cite{3GPPLatency}, each of the elements contributing to the total delay, i.e $d_{DS}$, can be estimated as a function of Transmission Time Interval (TTI) and SR period $T_{SR}$, as seen in the evolution of Equation (\ref{eq:d_ds}). The processing delays is estimated as three TTI in Release 14 \cite{3GPPLatency}, while for 5GNR it is estimated as one TTI \cite{Joachim}. Since our aim is to study the use of dynamic scheduling for haptic communications, and since we are referring to haptic radio delay, we consider the one TTI processing delay.

\subsection{Semi-Persistent Scheduling}

In latency critical applications, such as haptic communication that is the focus of this paper, repeating scheduling request procedure for every packets, as in dynamic scheduling, might not guarantee the application requirements. Also, depending on the pattern of data traffic, dynamic scheduling might introduce significant additional overhead. One available solution for reducing scheduling latency is static allocation of UL grant with a given periodicity to the traffic. This mechanism, referred to in the literature as Semi-Persistent Scheduling (SPS), is an effective way to reduce scheduling latency of deterministic traffic like VoIP, but it will have inverse effect on the spectral efficiency in the case of bursty traffic due to un-used allocated resource in non-burst periods. Also, SPS might fail to transmit all packets during burst periods\cite{SPS2012}. 

In SPS, after initialization, i.e. after the first UL grant, the radio access latency for the rest of UL transmissions can be expressed by $d_{SPS}$, according to Equation (\ref{eq:d_SPS}).

\begin{equation}
\begin{aligned}
\label{eq:d_SPS}
d_{SPS} \leq &T_{pg}+T_{RRB(UE)}+T_{proc(UE)} \\
&+T_{tx(UE \rightarrow eNB)}+T_{proc(eNB)} \\
=&T_{pg}+TTI+TTI+TTI+TTI
=T_{pg}+4TTI \text{,}
\end{aligned}
\end{equation}

where $T_{pg}$ is the period of the pre-allocated UL grant on Physical Downlink Control Channel (PDCCH), and $T_{RRB(UE)}$ denotes the time required by the UE to read the received Resource Block (RB) from eNB on PDCCH. Other notations are the same as in Equation (\ref{eq:d_ds}). Similar to calculation of $d_{DS}$, the estimated values of each delay component are extracted from \cite{3GPPLatency}.

\subsection{Soft-Resource Reservation}

To improve the spectral efficiency of SPS for bursty traffic, Soft-Resource Reservation (SRR) \cite{Massimo} has been proposed. In this scheduling mechanism, the UL grant is reserved until the UE has data to transmit, otherwise, the UL grant is released, and for further transmission the UE needs to send another SR. Therefore, UL radio access latency during burst period follows Equation (\ref{eq:d_SPS}) and during non-burst period follows Equation (\ref{eq:d_ds}).

\subsection{Fast UL access}

An enhanced version of SPS in order to accommodate the needs of URLLC communication, is proposed as the fast UL access (FA) \cite{3GPPLatency}. In the fast UL access, using puncturing mechanism, the UE has opportunity to transmit UL data every TTI after the data is ready to transmit. This scheduling mechanism makes UL radio access delay the same as DL radio access delay (this scheduling mechanism is also further discussed in \cite{Joachim}). Therefore, with fast UL access, latency critical application user can send UL packet immediately after packet generation on the next available TTI without getting permission from eNB. The maximum UL radio access latency, $d_{FA}$, could, therefore, be calculated as in Equation (\ref{eq:d_FA}).

\begin{equation}
\begin{aligned}
\label{eq:d_FA}
d_{FA} \leq &TTI+T_{proc(UE)}+T_{tx(UE \rightarrow eNB)}+T_{proc(eNB)} \\
=&TTI+TTI+TTI+TTI=4 TTI \text{.}
\end{aligned}
\end{equation}

Delay components of Equation (\ref{eq:d_FA}) is similar to Equation (\ref{eq:d_SPS}), except that $T_{pg}$ for fast UL access is equal to the TTI, since latency critical user can potentially transmit in each TTI without getting permission. Also, in fast UL access, there is no need to read the PDCCH before UL transmission and thus, $T_{RRB(UE)}$ equals zero.

\section{Delay Analysis for Leftover traffic}
\label{sec:SNC}
\subsection{System Model}
\begin{figure}
	\centering
	\includegraphics[width=6.5cm]{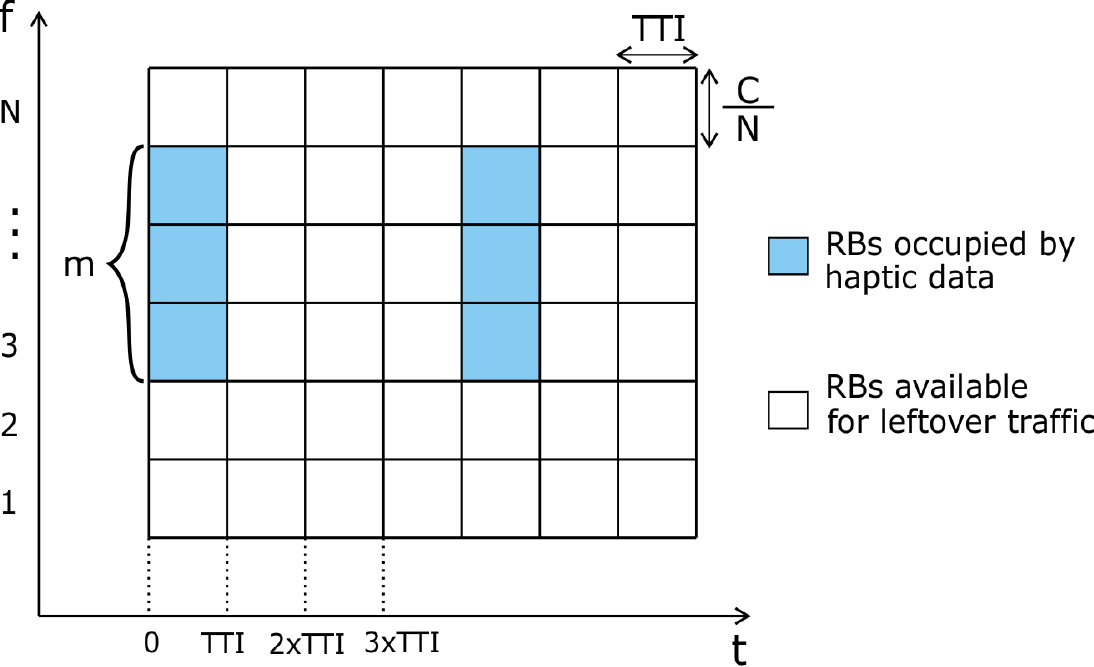}
	\caption{System Model\vspace{-5mm}}
	\label{fig:sys model}
\end{figure}

We consider a constant rate multi-channel server consisting of
 $N$ frequency channels with total transmission rate of $C$. Transmission rate of each channel is considered as $C/N$ as depicted in Figure \ref{fig:sys model}. All channels are synchronized in time and each square represents one Resource Block (RB) of time $\times$ frequency. The time unit for allocating RBs is one Transmission Time Interval (TTI). The coloured blocks are occupied by haptic data and the blank blocks will remain for use of leftover traffic. Since haptic data requires ultra-low latency, we assume that for transmission of each packet, only one block is occupied in time and it will spread in frequency to fulfil the resource block it needs. Denoting the amount of time $\times$ frequency resource needed for haptic data by $\rho$, the number of frequency blocks occupied by haptic data, denoted by $m$, is computed by the following formula:
 
 \begin{equation}
 \label{eq:m}
 m=\ceil{\frac{N \times \rho}{C\times TTI}} \text{ .}
 \end{equation}

After occupation of the coloured blocks by haptic data (i.e. $m$ blocks for each packet transmission opportunity), we will use Stochastic Network Calculus (SNC) to compute the service remained for the leftover traffic and the resulting stochastic delay bound.

\subsection{Methodology} 

The Stochastic Network Calculus (SNC) \cite{SNC} is a mathematical method being used for analyzing delay and backlog in a wide class of stochastic traffic and service models. In this paper, we will apply SNC to analyze the effect of using different UL scheduling mechanisms, for haptic communication, on the delay of the remainder of traffic, i.e. leftover service delay. Here, we will provide some definitions, from \cite{Jiang2006}, in the SNC that will be used in the next subsection for modeling and analysis.

\theoremstyle{definition}
\newtheorem{definition}{Definition}
\begin{definition}\label{def:SAC}
	\textit{Stochastic Arrival Curve (SAC):} The virtual backlog centric (v.b.c) SAC $\alpha$ bounded by function $f$ is defined for a flow if for all $x \geq 0$ and all $0 \leq \tau \leq t$ there holds:
	\begin{equation}
	P\{ \sup\limits_{0 \leq \tau \leq t} [ A(\tau,t)-\alpha(t-\tau) ] \geq x\}\leq f(x) \text{ .}
	\end{equation}
	where $A(\tau,t)$ is the cumulative arrival traffic during the time $(\tau,t]$.
	
\end{definition}

\begin{definition}\label{def:SSC}
	\textit{Stochastic Service Curve (SSC):} The weak SSC $\beta$ bounded by function $g$ is defined for a server if for all $x\geq 0$ and $t \geq 0$ there holds:
	\begin{equation}
	\begin{aligned}
	&P\{A \otimes \beta (t) - A^{\star} (t) > x\} \leq g(x) \\
	&A \otimes \beta (t) =
	\inf\limits_{0\leq{\tau}\leq{t}}\{A(\tau)+\beta(t-\tau)\} \text{ ,}
	\end{aligned}
	\end{equation}
	where $A^{\star}(t)$ is cumulative departure traffic during the time $(0,t]$ and $\otimes$ is the min-plus convolution.
\end{definition}

\newtheorem{theorem}{Theorem}
\begin{theorem}
	\textit{Stochastic Delay Bound :}
	The stochastic delay $D(t)$ of the arriving flow in a queuing system with SAC and SSC as defined in the Definitions \ref{def:SAC} and \ref{def:SSC}, respectively, is upper bounded as:
	\begin{equation}
	\label{eq:upper delay}
	P\{D(t) > h(\alpha + x,\beta)\}\leq f \otimes g(x) \text{ ,}
	\end{equation}
	where
	\begin{equation}
	\label{eq:horizontal}
	h(\alpha + x,\beta)=\sup\limits_{0\leq{\tau}\leq{t}}\{\inf\{s \geq 0:\alpha(\tau)+x\leq\beta(\tau+s)\}\}
	\end{equation}
	$h(\alpha + x,\beta)$ is the maximum horizontal distance between $\beta$ and $\alpha+x$.
\end{theorem}

\subsection{Modeling and Analysis}
In this part, we model the upper bound delay of leftover traffic, which refers to the remainder of traffic, after scheduling the haptic communication traffic. We formulate this problem similar to the case of cognitive radio with primary users, being haptic communication flows, and secondary users, being the remainder of traffic \cite{Cognitive}. Since in SNC, a service curve is defined as a lower bound service provided to the traffic, in order to compute the leftover service curve, we should consider the worst case time period $(\tau,t]$ $0 \leq \tau \leq t$. The leftover service would be greater than its worst case during $(\tau,t]$:

\begin{equation}
\begin{aligned}
\label{eq:Service LO}
S_{lo} \geq \frac{(N-m)C}{N}(t-\tau) + \frac{mC}{N} (t-\tau-R TTI-2TTI)^+ \text{,}
\end{aligned}
\end{equation}

where $S_{lo}$ denotes the guaranteed leftover service, $a^+=max\{a,0\}$, and $R$ is the number of RBs occupied by haptic traffic during the worst case time period $(\tau,t]$. The first sentence in the right hand side is the service available for leftover traffic in $(N-m)$ sub-channels that weren't used by haptic traffic, thus all period of $(\tau,t]$ can be used by the leftover traffic. The second sentence represents the leftover service in $m$ sub-channels that are occupied by haptic traffic, thus, the $R$ time periods have been subtracted from $(\tau,t]$ time period. Also, two TTIs are excluded since in the worst case, $\tau$ might be just after TTI start and $t$ might be just before TTI end.

The right hand side of Equation (\ref{eq:Service LO}) is defined as leftover Stochastic Service Curve (SSC) $\beta_{lo}$ which is a function of time period $(\tau,t]$:

\begin{equation}
\label{eq:beta}
\beta_{lo}(t-\tau) \triangleq \frac{(N-m)C}{N}(t-\tau) + \frac{mC}{N} (t-\tau-R TTI-2TTI) \text{.}
\end{equation}

Note that we remove the function $a^+$ for simplicity and since $a^+ \geq a$ which still satisfies Inequality (\ref{eq:Service LO}). We consider Stochastic Arrival Curve (SAC) of leftover traffic as compound Poisson Levy process which is the summation of all packet lengths arrived till time $t$, with packet lengths of i.i.d. random variables with distribution of $exp(\sigma)$ and number of packets with distribution of independent Poisson process with parameter $\lambda$. The corresponding SAC would be \cite{Cognitive}:

\begin{equation}
\begin{aligned}
\label{eq:SAC}
\alpha_{lo}(t)=\frac{\lambda t}{\theta}(e^{\theta \sigma}-1) \quad , \quad f(x)=e^{-\theta x} \text{,}
\end{aligned}
\end{equation}

where $\theta \geq 0$ (free parameter). Let $d_0=h(\alpha_{lo}(t)+x, \beta_{lo})$, according to the definition in Equation (\ref{eq:horizontal}), we can compute $d_0$ by the equation $\beta_{lo}(d_0)=x$. We assume the communication channel will serve all packets of leftover traffic (the communication channel is error free). Hence, $g(x)$ in Equation (\ref{eq:upper delay}) equals zero and $f \otimes g(x)=f(x)$. By substituting into Equation (\ref{eq:upper delay}), we can get:

\begin{equation}
\begin{aligned}
\label{eq:d_lo}
P\{d_{lo} > d_0\} \leq e^{-\theta \beta_{lo}(d_0)}\text{,}
\end{aligned}
\end{equation}

where $d_{lo}$ denotes leftover traffic delay. One can compute the upper bound of $d_{lo}$ with outage probability $\epsilon$ (i.e. $d_0$) when the right hand side of the Equation (\ref{eq:d_lo}) is equal to $1-\epsilon$. Note that to have a stable queue, the arrival rate must be less than or equal to the service rate, thus, $\lambda (e^{\theta \sigma}-1)/ \theta$ must be less than the coefficient of $(t-\tau)$ in $\beta_{lo}(t-\tau)$. Also note that we can get the tightest upper bound with the maximum $\theta$ that satisfies the stability condition.

\section{Effect of Haptic UL Scheduling on Leftover Traffic}
\label{sec:problem}

\begin{figure}
	\centering
	\includegraphics[width=\linewidth]{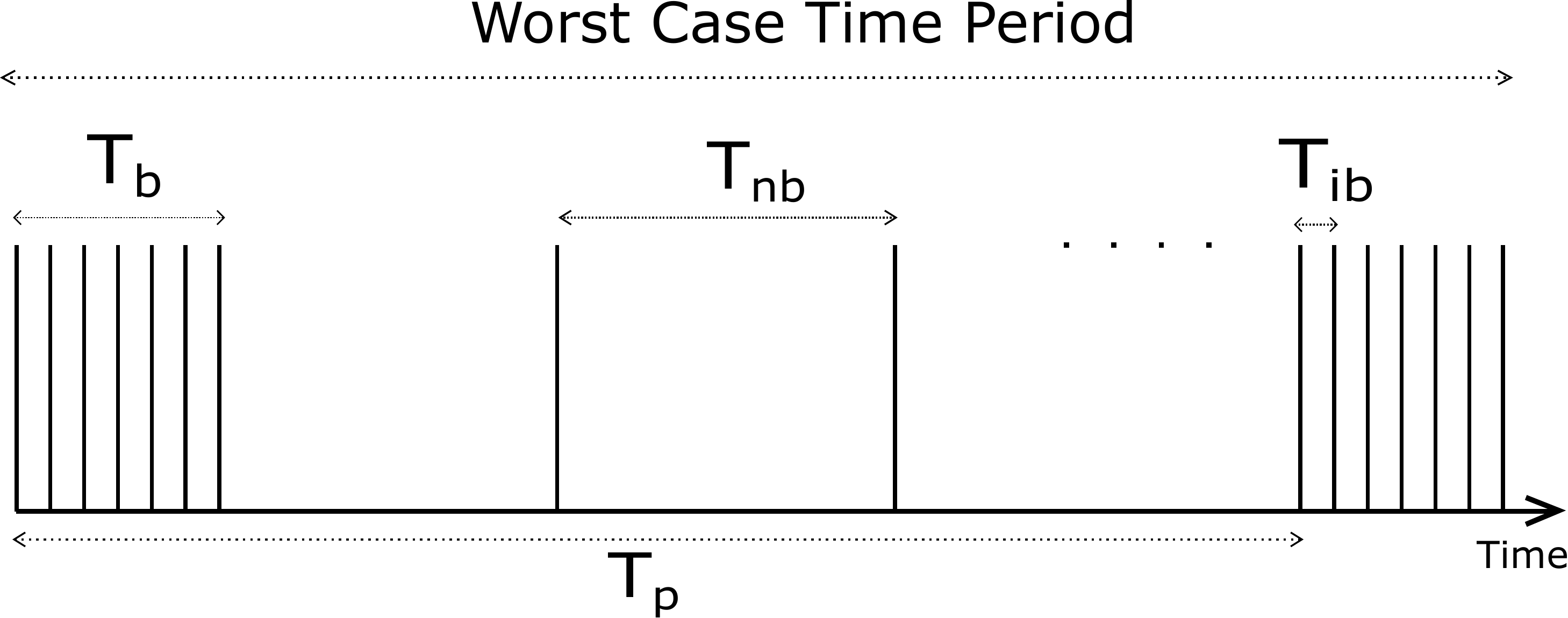}
	\caption{Model of haptic traffic packet arrivals. In this figure, $T_p$ is the duration of one traffic period, $T_b$ is duration of each burst, $T_{nb}$ is packet inter arrival time in non-burst periods, and $T_{ib}$ is packet inter arrival time during burst periods.\vspace{-5mm}}
	\label{fig:Haptic Traffic}
\end{figure}

To model the effect of UL scheduling mechanism used for haptic communications, as a case of URLLC, on the leftover service and delay, firstly, we need to consider a model for haptic traffic packet arrivals. Based on experiments in \cite{Xiao2017}, we consider periodic bursty traffic for haptic data as depicted in Figure \ref{fig:Haptic Traffic}. Bursts arrive with period of $T_{p}$, their duration is $T_b$, and the inter arrival time during the burst period is $T_{ib}$. Also, during non-burst period, haptic data arrives every $T_{nb}$. Note that in the worst case time period, there would be an excess burst other than the bursts inside the periodic traffic like the one shown at the end of $T_p$ in Figure \ref{fig:Haptic Traffic}.

In our analysis, we assume that haptic data is a part of the global closed control loop (such as position, force and velocity in teleoperation system) and thus, when there is an opportunity to start transmission to eNB, only the latest version of haptic data is meaningful to be transmitted (as the most up-to-date position, force, etc.) and the previous un-transmitted data will be discarded.

With the aforementioned assumptions, we try to compute the upper bound delay of leftover traffic for each of the UL scheduling mechanisms described in section \ref{sec:delay}, using the model detailed in Section \ref{sec:SNC}. In fact, we only need to find $\beta_{lo}(t-\tau)$ in Equation (\ref{eq:beta}) and then use it to find $d_0$ in Inequality (\ref{eq:d_lo}).

\subsection{Dynamic Scheduling}
\label{sec:problem_DS}
If haptic packets inter arrival time during burst periods satisfies Inequality (\ref{eq:Tib}), then no haptic data packet would be discarded by the UE. In other words, this inequality ensures that the UE can receive a grant for the arrived packet before the arrival of any new packet.

\begin{equation}
\label{eq:Tib}
T_{ib} > T_{SR}+T_{tx(UE \rightarrow eNB)}+T_{proc(eNB)}+T_{tx(eNB \rightarrow UE)} \text{  ,}
\end{equation}

In Inequality (\ref{eq:Tib}), the right hand side is the first four sentences of Equation (\ref{eq:d_ds}), i.e. the delay before processing the grant and encoding the packet by the UE. Therefore, the SSC in this case would be,

\begin{equation}
\label{eq:beta_DS}
\begin{aligned}
&\beta_{lo-DS}(t-\tau) \triangleq \frac{(N-m)C}{N}(t-\tau) +  \\
&\frac{mC}{N} (t-\tau- n_p R_p TTI-R_b TTI-2TTI) \text{ ,} \\
&n_p=\floor{\frac{t-\tau}{T_p}} \quad , \quad R_b=\floor{\frac{T_b}{T_{ib}}}\\
&R_{nb}=\floor{\frac{T_p-T_b}{T_{nb}}} \quad , \quad
R_p=R_b+R_{nb} \text{ ,}
\end{aligned}
\end{equation}

where $n_p$ is the number of whole traffic periods within $t-\tau$. $R_b$, $R_{nb}$, and $R_p$ are the number of transmitted haptic packets during each burst period $T_b$, each non-burst period $(T_p-T_b)$, and each traffic period $T_p$, respectively. As we discussed before, each haptic packet consumes one TTI in time domain but it spreads in frequency domain to achieve lower latency.

If $T_{ib}$ doesn't satisfy the Inequality (\ref{eq:Tib}), but is still greater than twice of the right hand side of the Inequality (\ref{eq:Tib}), half of the haptic packets during burst period would noti have the chance to be transmitted and therefore $R_b$ in Equation (\ref{eq:beta_DS}) would be $R_b=\floor{\frac{T_b}{2T_{ib}}}$. For lower $T_{ib}$, $R_b$ changes accordingly in a same way.

\subsection{Semi-Persistent Scheduling}

In the SPS, a periodic grant is allocated for the haptic UE regardless of whether it has any packet to transmit on this or not. Assuming the period of the allocated grant is $T_{pg}$, the SSC would be,

\begin{equation}
\label{eq:beta_SPS}
\begin{aligned}
&\beta_{lo-SPS}(t-\tau) \triangleq \frac{(N-m)C}{N}(t-\tau) +  \\
&\frac{mC}{N} (t-\tau-n_p R_{pg} TTI-R_{bg} TTI-2TTI)  \\
&R_{pg}=\floor{\frac{T_p}{T_{pg}}} \quad , \quad R_{bg}=\floor{\frac{T_b}{T_{pg}}} \text{  , }
\end{aligned}
\end{equation}

where $R_{pg}$ and $R_{bg} $ are the number of allocated grants during each haptic traffic period and each haptic burst periods, respectively. In SPS, however, some allocated resources might be unused during non-burst periods and also some haptic packets might be discarded during burst periods.

\subsection{Soft-Resource Reservation}

In the SRR, during burst periods the UE keeps the periodic allocated grant (which has the period of $T_{pg}$) and transmit on the reserved RBs, but during non-burst periods, it will send SR and waits for the permission to transmit. Therefore, from the leftover RBs point of view, it is the same as SPS during bursts and the same as DS outside the bursts. Therefore, its SSC would be:

\begin{equation}
\label{eq:beta_SRR}
\begin{aligned}
&\beta_{lo-SRR}(t-\tau) \triangleq \frac{(N-m)C}{N}(t-\tau) +  \\
&\frac{mC}{N} (t-\tau- n_p (R_{bg}+R_{nb}) TTI-R_{bg} TTI-2TTI) \text{ ,}
\end{aligned}
\end{equation}

where $n_p$, $R_{bg}$, and $R_{nb}$ are the same as previously defined in Equation (\ref{eq:beta_DS}) and (\ref{eq:beta_SPS}).

\subsection{Fast UL access}

In fast UL access, if haptic packet inter arrival time is greater than TTI, no packet would be discarded and every packet would have the chance to be transmitted. In that case, SSC would be the same as DS in Equation (\ref{eq:beta_DS}). However, haptic packets would experience much less delay since there is no need for them to send SR and wait for the UL grant before starting transmission. If $T_{ib} <$TTI, some packets might be discarded and $R_b$ would differ accordingly as discussed in section \ref{sec:problem_DS}.

\section{Numerical Results}
\label{sec:simulation}

This section presents the numerical results of the model described earlier. We assume haptic communication services use one of the four discussed UL scheduling and we see the impact of this on the remaining service rate, used for the leftover traffic. It is important to note that the UL scheduling delay for the leftover traffic is not included in the leftover delay plots here, and the presented delay is just the impact of remainder of service.

These analyses are performed using the parameters in Table \ref{table:parameters}.
%
%
 The haptic traffic model is based on the presented results in \cite{Xiao2017}, this traffic pattern is transmitted data after haptic lossy encoder has been applied to the haptic data. Using the assumptions in Table \ref{table:parameters} and Equations (\ref{eq:d_ds})-(\ref{eq:d_FA}), maximum radio access delay resulted by UL scheduling for haptic packets, are: $7$TTI, for DS, $4$TTI, for FA, and $14$TTI for SPS and SRR (during the burst period).

Main performance indicators studied in this paper are as follows. First, we study how using each of the scheduling would affect the packet drop rate of haptic communications. Dropping haptic packets here is based on the assumption that a haptic packet will only be transmitted if it can be scheduled upon its arrival and it will not be buffered (the queue length is zero). This assumption relies on the fact that haptic data refers to physical attributes of position, velocity, etc, and when an update value of these attributes become available, the old value is not valid anymore. In addition, and due to the fact that the haptic data is encoded data, and every single packet should be transmitted for delivering high quality haptic communication, we consider only zero-packet-drop regions as our operational region. Second performance metric we look into is the \textit{remainder of service}, referring to what service rate remains for the leftover traffic. We also examine the upper bound delay that can be offered to the leftover traffic, using this remainder of service.

\begin{table}
	\caption{Simulation parameters}
	
	\begin{center}
		\scalebox{0.7}{
		\begin{tabular}{|c|c|}
            \hline
            parameter & value \\
            \hline
            \hline
            leftover traffic & \cite{traffic-model} \\
            \hline
            $\lambda $ & 4~packets/sec \\
            $\sigma$ & 1500~bytes/sec \\
            \hline
            \hline
            Communication Channel & \cite{Joachim},\cite{3GPPLatency}\\
            \hline
            TTI  & 0.125, 0.25, 0.5, 1~ms \\
            $T_{pg}$ & $10\times$TTI\\
            $T_{SR}$ & TTI\\
            \hline
            \hline
            SNC parameters & \cite{Cognitive}\\
            \hline
            $\epsilon$  & $10^{-5}$ \\
            RB  & $10^{-3}$ \\
            $N$  & 10 \\
            \hline
            \hline
            Traffic timing parameters & \cite{Massimo} \\
            \hline
            $T_{ib}$ & 1-3~ms\\
            $T_{nb}$ & 50~ms\\
            $T_{b}$ & 200~ms\\
            $T_{p}$ & 1000~ms\\
            \hline
		\end{tabular}}
\vspace{-5mm}
	\end{center}
	\label{table:parameters}
\end{table}

\begin{figure*}[!h]
	\label{fig:UL_TTI}
	\centering
	\subfigure[]{
		\includegraphics[width=.62\columnwidth,height=3cm]{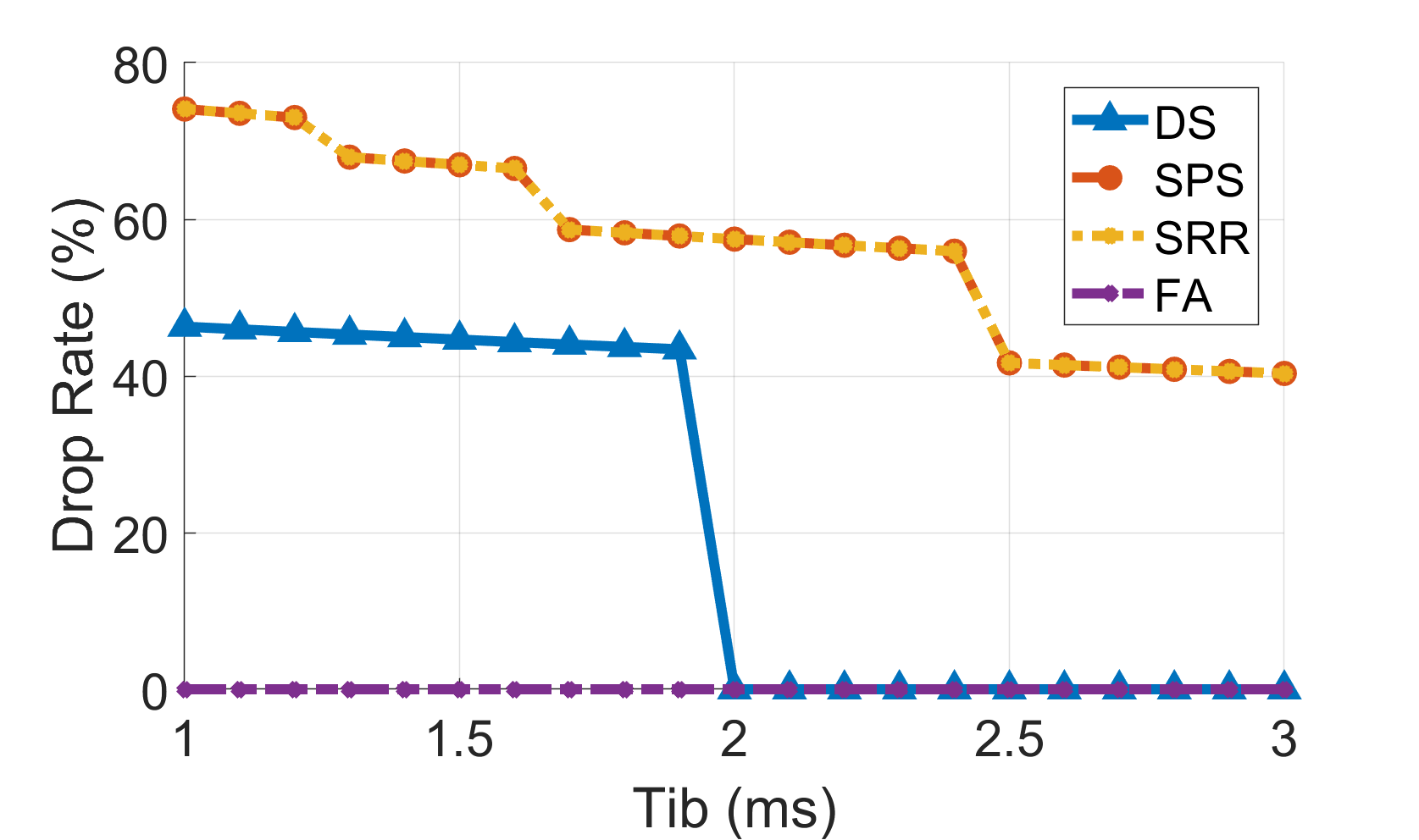}
		\label{fig:Dr_vs_Tib}
	}
	\subfigure[]{
		\includegraphics[width=.62\columnwidth,height=3cm]{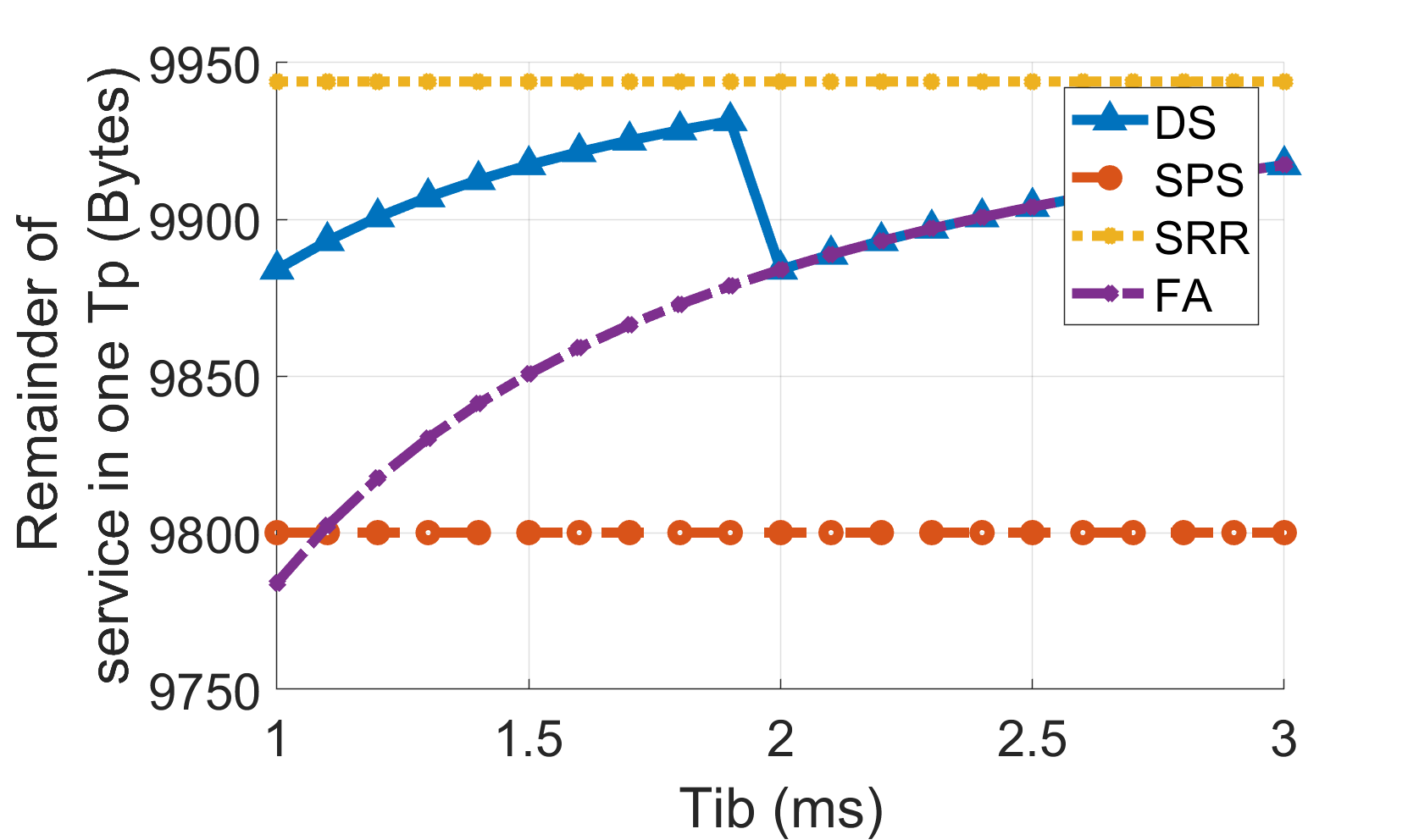}
		\label{fig:S_vs_Tib}
	}
	\subfigure[]{
	\includegraphics[width=.62\columnwidth,height=3cm]{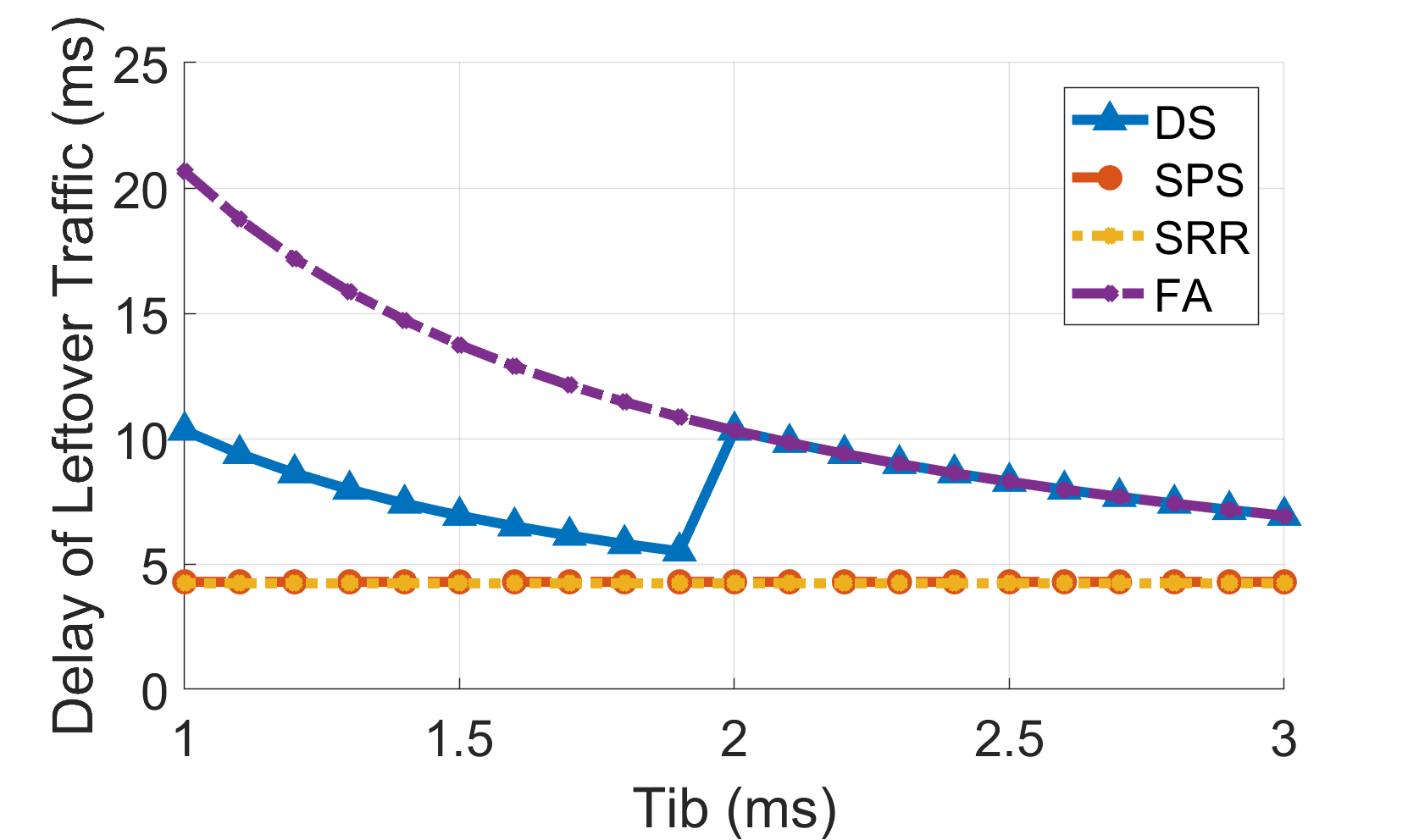}
	\label{fig:dlo_vs_Tib}
	}
\caption{(a) Haptic packet drop rate, (b) remainder of service, (c) leftover traffic upper bound delay, Vs. packet inter arrival time of haptic traffic during burst periods ($T_{ib}$), using different UL scheduling schemes with TTI = 0.5 ms.\vspace{-5mm}}
\end{figure*}

Figure \ref{fig:Dr_vs_Tib} shows the haptic packet drop rate for different UL scheduling mechanisms when TTI = $0.5$~ms and $T_{pg}$ = $10\times$TTI. It can be seen that using SPS and SRR cannot offer zero packet drop rate in the observed range of $T_{ib}$, thus, SPS and SRR are not good candidates with the above mentioned timing (i.e. TTI and $T_{pg}$). In addition, for DS, as soon as arriving haptic data has higher frequency than one packet per $2$~ms, packet loss is introduced. It can be concluded that for $T_{ib}$ below $2$~ms, only FA can be used as UL scheduling mechanism for haptic communication, and for $T_{ib}$ between $2$ and $3$~ms, both FA and DS can be used.

Figure \ref{fig:S_vs_Tib} shows the remainder of service (to be used by the leftover traffic) in one period of haptic traffic vs packet inter arrival time of haptic data during burst periods, i.e. $T_{ib}$ at TTI = 0.5~ms ($T_{pg}$ is still 5~ms). Observing from this figure, the remainder of service when SRR is used is higher than when SPS is used, since SRR releases the grant in non-burst periods. Another important observation is when SPS and SRR are used, the leftover service is independent of $T_{ib}$, since the pre-allocated periodic grant with the constant period of $T_{pg}$ is occupied by the haptic traffic regardless of its data generation frequency.

Further observation from Figure \ref{fig:S_vs_Tib} shows for $T_{ib}$ larger than 2~ms, the remainder of service behaves very close when either of DS and FA are used; this is the region in which both DS and FA offer zero packet drop. Also note that by using DS or FA, the remainder of service increases with the increase in $T_{ib}$, since the larger $T_{ib}$, the fewer packets are generated by the haptic communication flow and thus, the more resource would remain for the leftover traffic. Given that the remainder of service does not show significant differences, deploying different UL scheduling techniques, we examine the upper bound delay offered by this service closely.

Observation from Figure \ref{fig:dlo_vs_Tib} shows that the upper bound latency is affected more significantly. In other words, if the remainder of service is used for traffic with latency constraint, deploying any of the UL scheduling will have a major impact on its performance, while this is not the case for no delay constraint traffic. Figure \ref{fig:dlo_vs_Tib} shows the upper bound delay of leftover traffic, computed based on Equation (\ref{eq:d_lo}). As we can see, for $T_{ib}$ between $2$ and $3$~ms, DS and FA perform similarly. In summary, results in Figures \ref{fig:Dr_vs_Tib}-\ref{fig:dlo_vs_Tib} confirm the superiority of FA UL scheduling mechanism for haptic communication at TTI = $0.5$~ms and also possibility of using DS for $T_{ib} \geq 2$~ms.

%
%

To see the impact of different TTI values, we also vary TTI from $0.125$~ms to $1$~ms, and plot upper bound delay for the leftover traffic and the packet drop rate of haptic communication flow for the four UL scheduling mechanisms (Figure 5). In general, lowering the TTI will lower the packet drop rate for haptic communication flow, while increasing the upper bound delay for the leftover traffic. For DS and FA, wherever the haptic drop rate is zero, the leftover delay is equal for all TTIs. The reason is that we assumed constant RB and $m$ will change according to the Equation (\ref{eq:m}). Since $m$ is inversely proportional to TTI, the effect of changing TTI is completely compensated (see Equation (\ref{eq:beta_DS})). However, we can use FA for all TTIs in the whole range of $T_{ib}$ since haptic packet drop rate is always zero, while we cannot use DS when TTI = $0.5$~ms for $T_{ib}<2$~ms and  when TTI = $1$~ms. Interestingly, we can use SPS and SRR when TTI = $0.125$~ms for $T_{ib}\geq 1.3$~ms as well as when TTI = $0.25$~ms for $T_{ib}\geq 2.5$~ms, but still the upper bound delay of the leftover traffic with those parameters are close to the upper bound delay of FA and DS. However, the maximum haptic packet delay would be $14$TTI for SPS ans SRR, while it would be $7$TTI for DS and $4$TTI for FA.

Note that for the TTIs and $T_{ib}$s that SPS and SRR are usable, both are almost the same in terms of the upper bound delay for the leftover traffic, but for example observing the results for $T_{pg}$ = $4\times$TTI at TTI = $0.125$~ms shows that SRR results in 25\% lower leftover delay than SPS (because of releasing the grant in non-burst periods).

%
%
\begin{figure*}[!h]
	\label{fig:UL}
	\centering
	\subfigure[]{
		\includegraphics[width=.7\columnwidth]{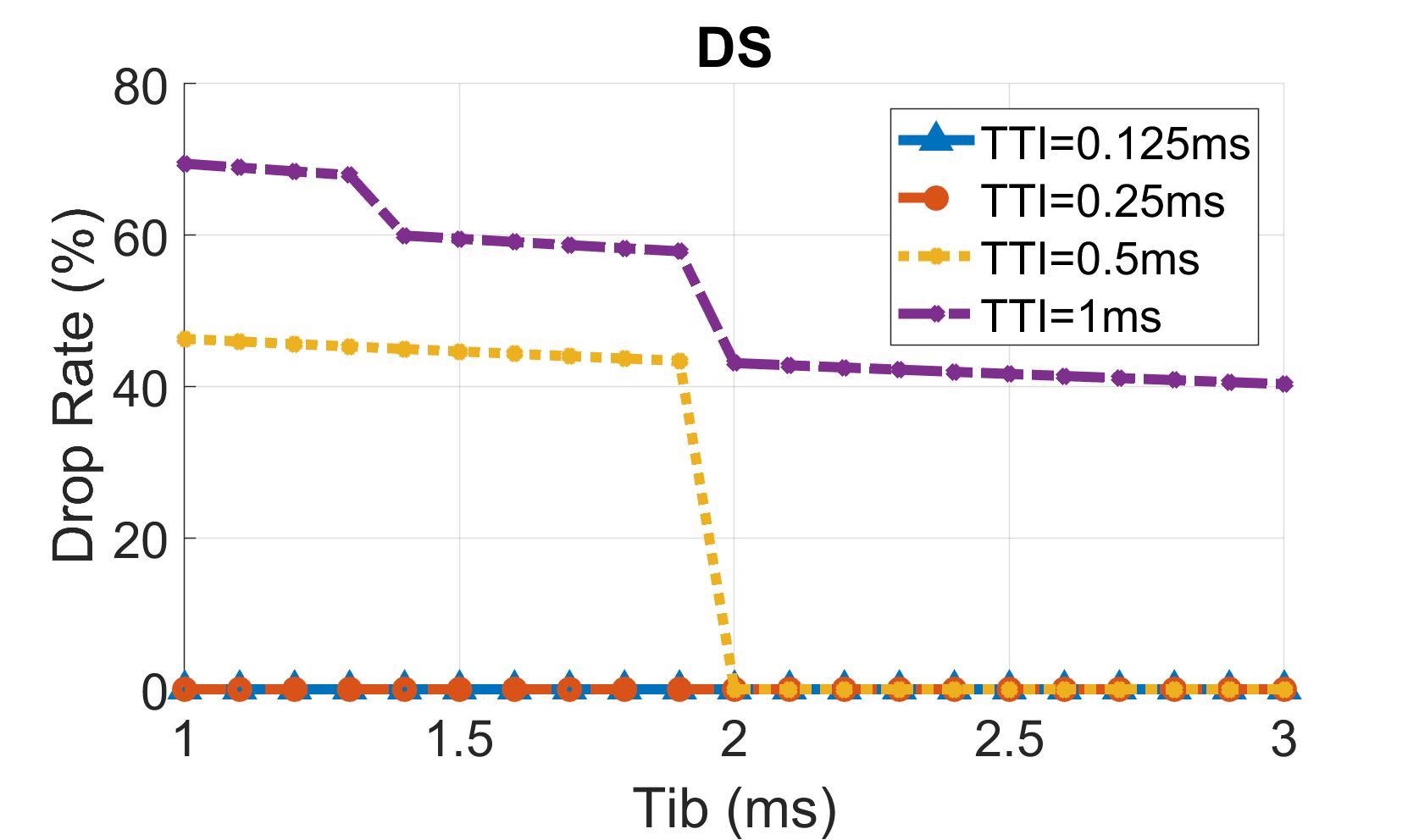}
		\label{fig:nrGroup}
	}
	\subfigure[]{
		\includegraphics[width=.7\columnwidth]{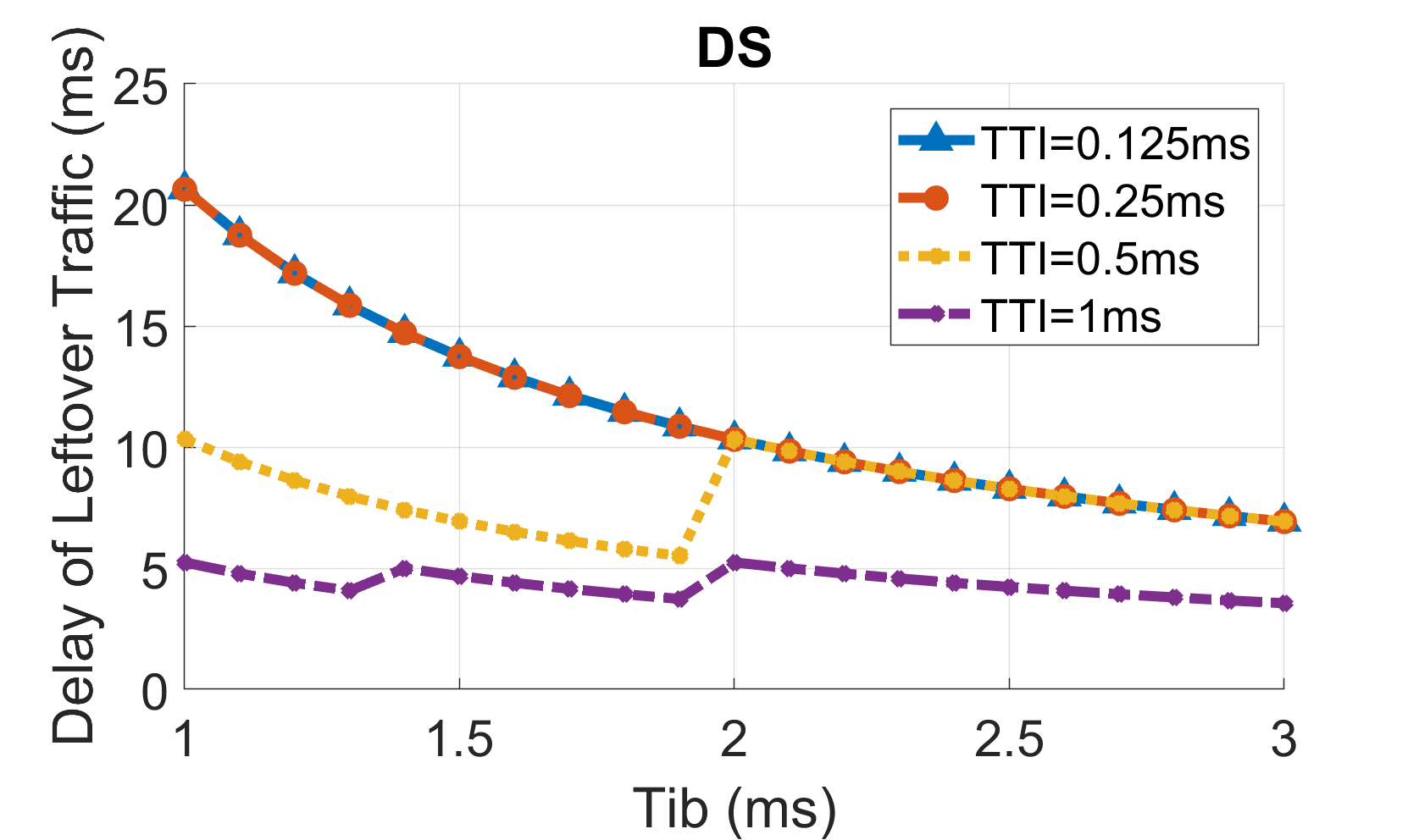}
		\label{fig:overallResult}
	}
\\
	\subfigure[]{
	\includegraphics[width=.7\columnwidth]{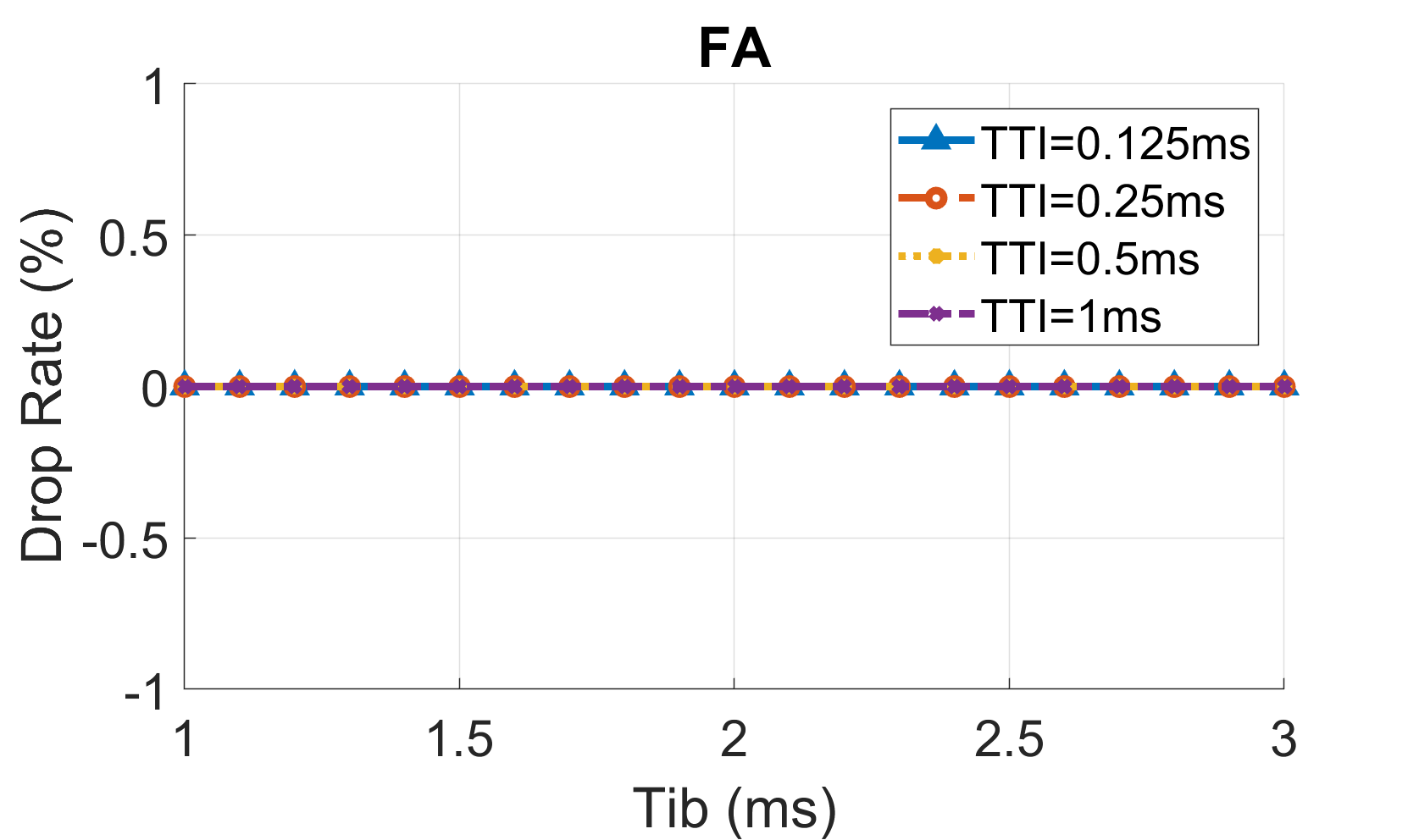}
	\label{fig:nrGroup}
	}
	\subfigure[]{
	\includegraphics[width=.7\columnwidth]{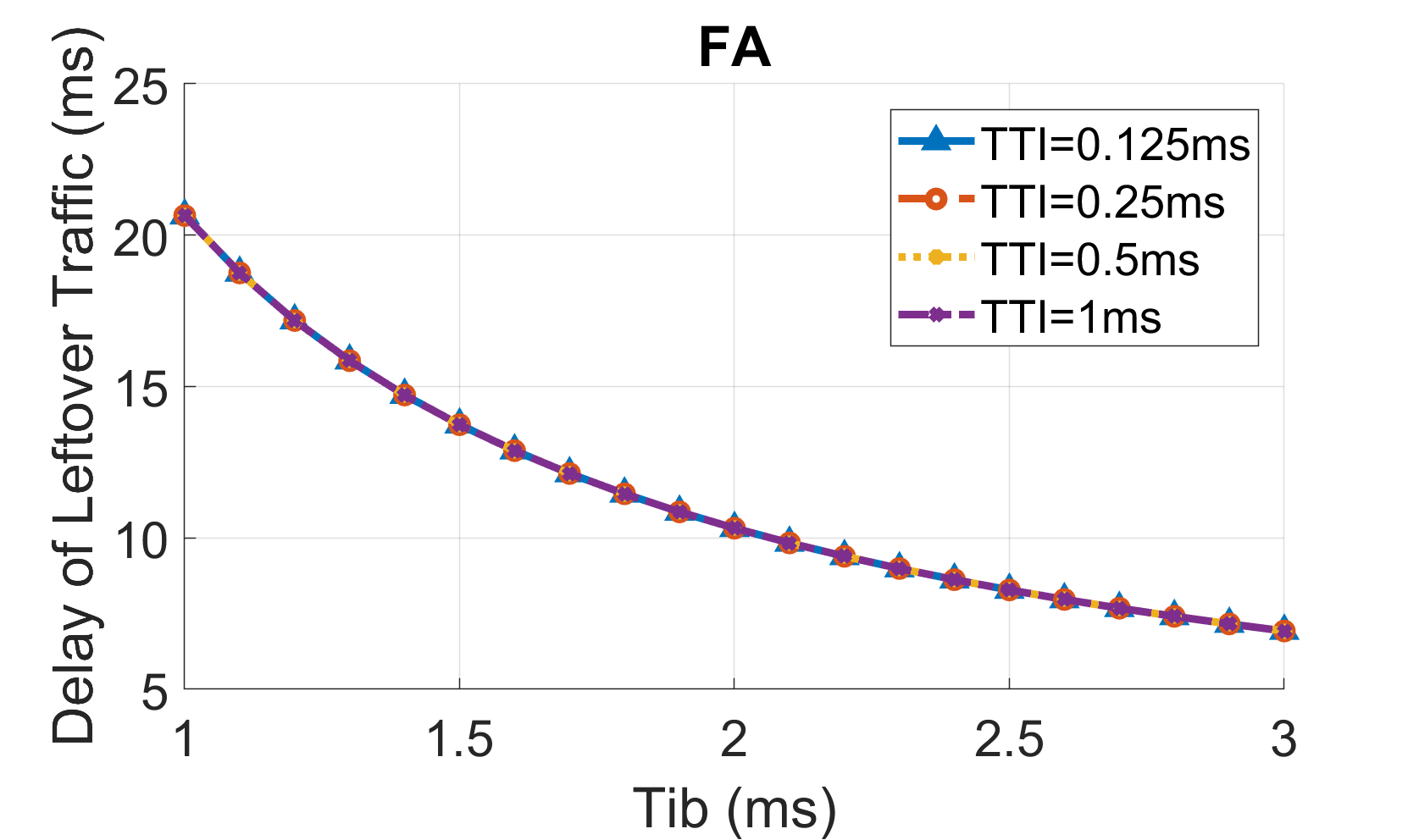}
	\label{fig:overallResult}
    }
\\
	\subfigure[]{
		\includegraphics[width=.7\columnwidth]{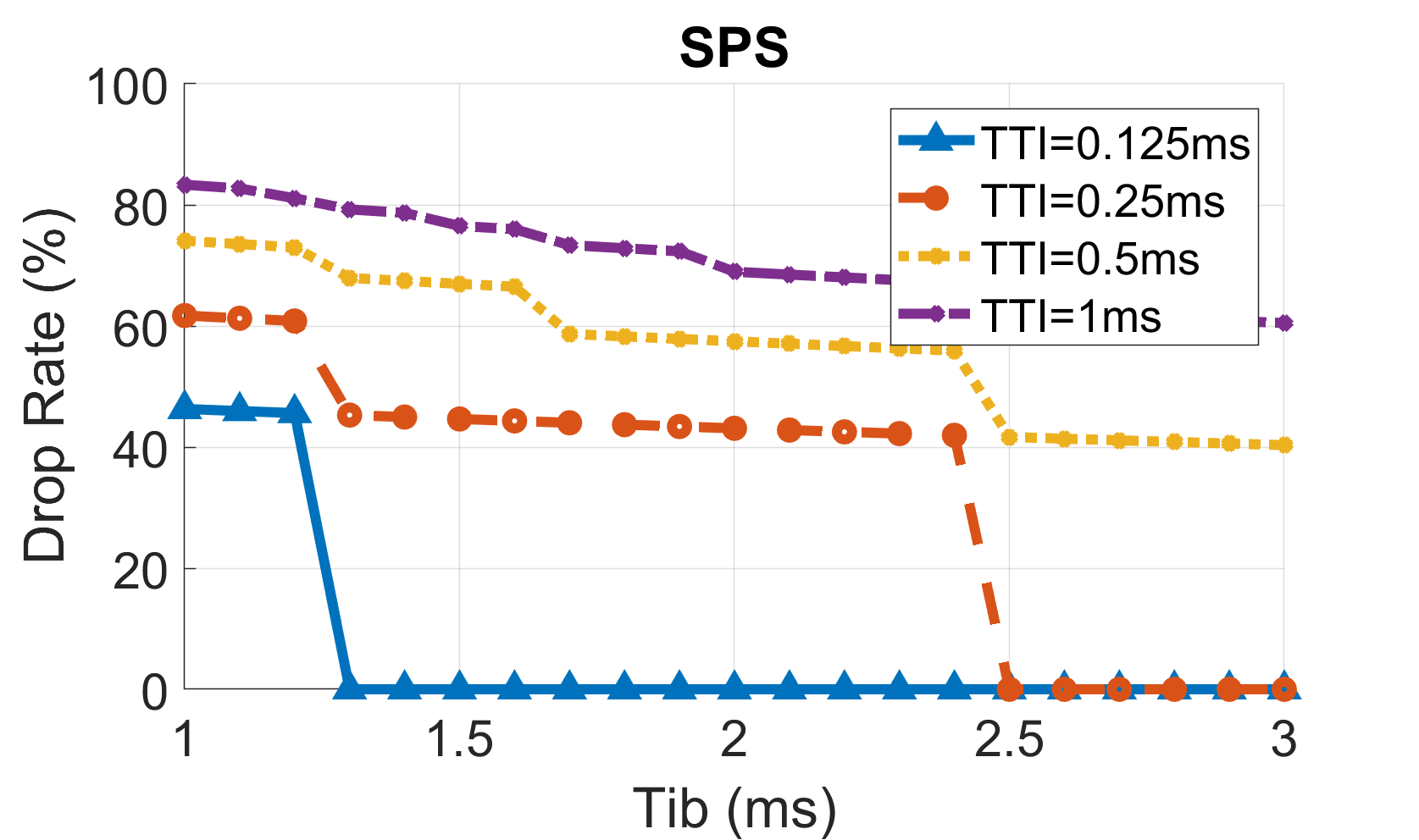}
		\label{fig:nrGroup}
	}
	\subfigure[]{
		\includegraphics[width=.7\columnwidth]{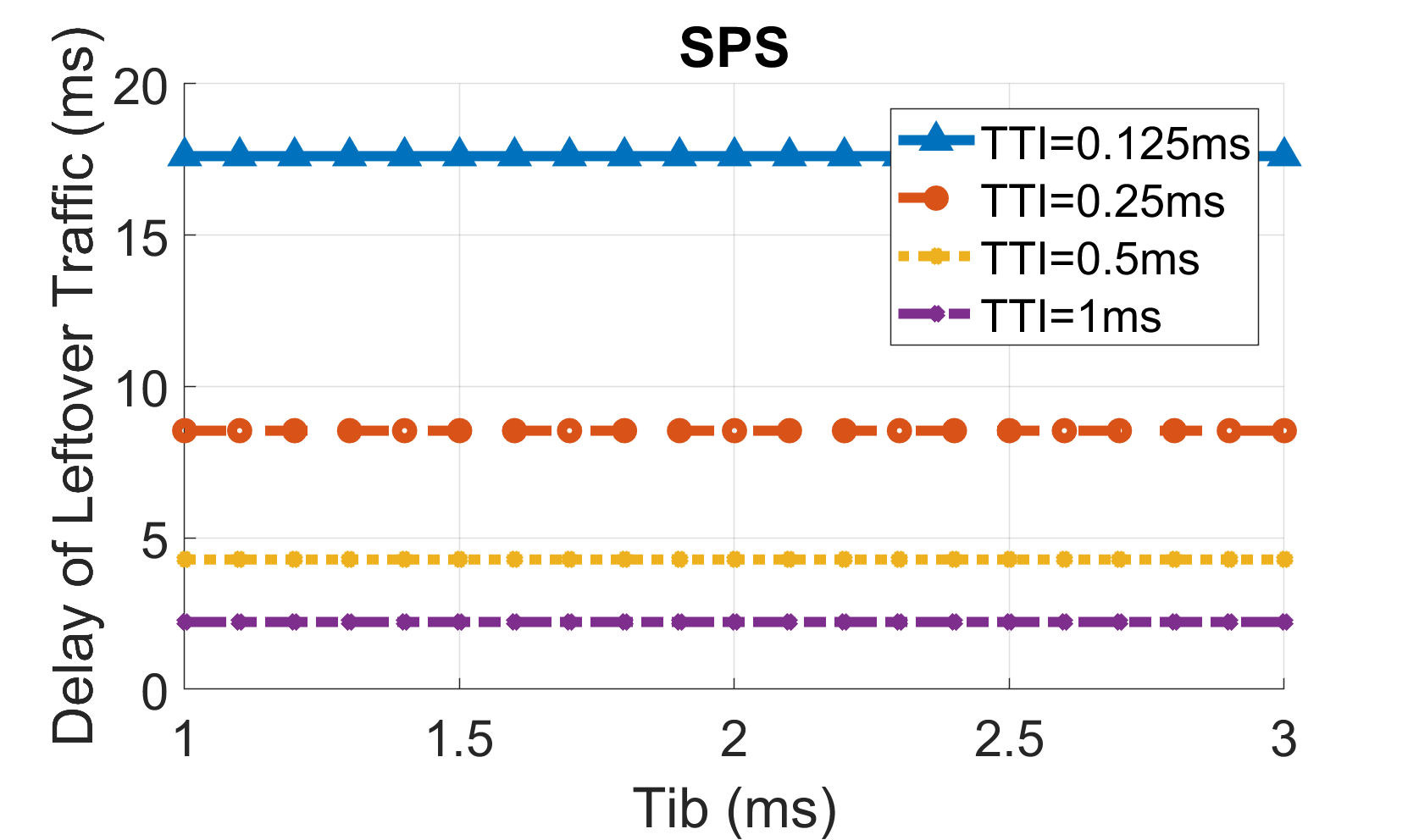}
		\label{fig:overallResult}
	}
\\
	\subfigure[]{
		\includegraphics[width=.7\columnwidth]{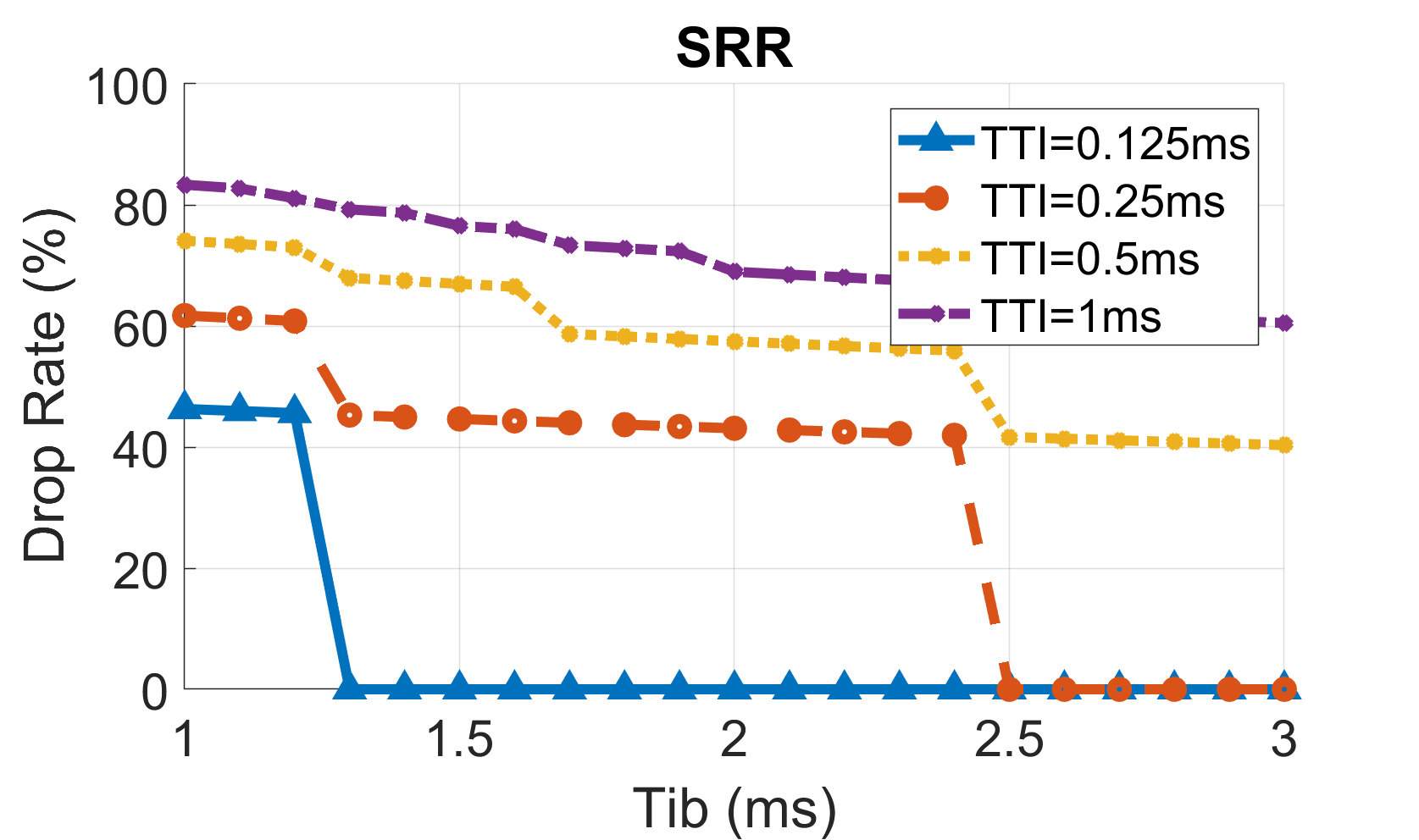}
		\label{fig:nrGroup}
	}
	\subfigure[]{
		\includegraphics[width=.7\columnwidth]{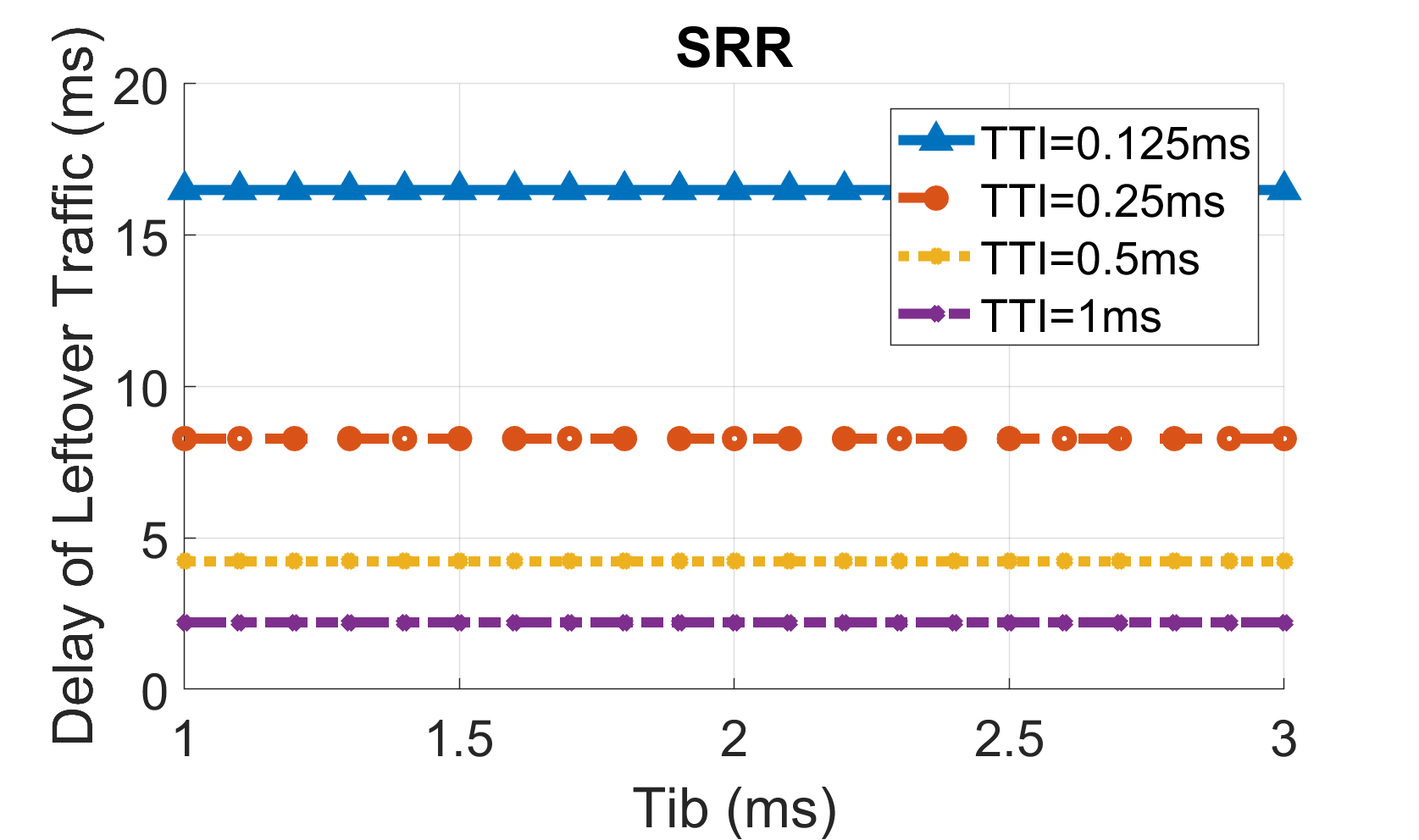}
		\label{fig:overallResult}
	}

	\caption{
	Haptic packet drop rate (left) and leftover delay (right) Vs. packet inter arrival time of haptic traffic during burst periods ($T_{ib}$).\vspace{-5mm}
	}
\end{figure*} 

\section{Concluding Remarks}
\label{sec:conclusions}

We have reviewed different UL scheduling mechanisms that can be used for haptic communication. Those include Dynamic Scheduling (DS), Semi-Persistent Scheduling (SPS), Soft-Resource Reservation (SRR), and Fast UL Access (FA). For each mechanism, we have modeled remaining service and delay for the leftover traffic after consuming the resources by haptic packets. Using the model, we obtained the regions that each UL scheduling mechanism can be used for communication of haptic packets. we observed that FA can be used for all TTIs and all typical inter arrival time of haptic packets during the burst period (based on the traffic model in \cite{traffic-model}) and also, this mechanism provides the lowest haptic packet radio access delay, while it has almost the same effect on the remainder of traffic (in terms of remainder service and upper bound of leftover delay) as other applicable UL scheduling mechanisms.

\begin{spacing}{0.88}
\bibliographystyle{ieeetr}
\bibliography{Globecomm_2018}
\end{spacing}
\end{document}